\RequirePackage{ifpdf}

\documentclass{PoS}
\usepackage[authoryear,square]{natbib}
\bibpunct{(}{)}{;}{a}{}{,}

\usepackage{epstopdf}

\title{HI galaxy simulations for the SKA: number counts and bias}

\ShortTitle{HI galaxy simulations}

\author{Mario G. Santos$^{1,2}$, David Alonso$^3$, Philip Bull$^4$, Marta Silva$^{5,1}$, Sahba Yahya$^1$\\
$^1$ Department of Physics, University of Western Cape, Cape Town 7535, South Africa\\
$^2$ SKA SA, 4rd Floor, The Park, Park Road, Pinelands, 7405, South Africa\\
$^3$Astrophysics, University of Oxford, DWB, Keble Road, Oxford OX1 3RH, UK \\
$^4$Institute of Theoretical Astrophysics, University of Oslo, PO Boks 1029 Blindern, 0315 Oslo, Norway\\
$^5$ CENTRA, Instituto Superior T\'{e}cnico, Universidade de Lisboa, Lisboa 1049-001, Portugal\\
E-mail: \email{mgrsantos@uwc.ac.za}}

\abstract{This chapter describes the assumed specifications and sensitivities for HI galaxy surveys with SKA1 and SKA2. It addresses the expected galaxy number densities based on available simulations as well as the clustering bias over the underlying dark matter. It is shown that a SKA1 HI galaxy survey should be able to find around $5\times 10^6$ galaxies over 5,000 deg$^2$ (up to $z\sim 0.8$), while SKA2 should find $\sim 10^9$ galaxies over 30,000 deg$^2$ (up to $z\sim 2.5$). The numbers presented here have been used throughout the cosmology chapters for forecasting.}

\FullConference{
Advancing Astrophysics with the Square Kilometre Array\\
June 8-13, 2014\\
Giardini Naxos, Italy}

\newcommand{\skipthis}[1]{}

\newcommand{\apjl}{ApJ}
\newcommand{\apjs}{ApJS}
\newcommand{\aap}{A \& A}

\newcommand{\be}{\begin{equation}}
\newcommand{\ee}{\end{equation}}
\newcommand{\bea}{\begin{eqnarray}}
\newcommand{\eea}{\end{eqnarray}}

\begin{document}

\section{Introduction}

Large scale structure surveys have so far relied on imaging of a large number of
galaxies at optical or near-infrared wavelengths combined with redshift information
to obtain the 3-dimensional galaxy distribution (e.g. DES, BOSS, Euclid). Hydrogen
is the most abundant baryon in the Universe, making it a prime candidate for a tracer
of the underlying dark matter distribution and an invaluable probe of the energy
content of the Universe and the evolution of large-scale structure.
In order to use hydrogen as a tracer at radio wavelengths, we need to rely on radio observations of the
neutral hydrogen (HI) 21cm line. With a rest frequency of 1420 MHz, telescopes probing
the sky between this and 250 MHz would be able to detect galaxies up to redshift 5.
Although we expect most galaxies to contain HI, since the spin-flip transition
responsible for the 21cm signal is highly forbidden, the emission is quite weak: at
$z=1.5$ most galaxies with a HI mass of $10^9\ \rm M_\odot$ will be
observed with a flux density of $\sim1\ \mu\rm Jy$. This implies that for low sensitivity
surveys, only the closest or HI-richest galaxies will be observed, and therefore
massive telescopes will be required in order to detect sufficient numbers of HI
galaxies at a level capable of providing competitive constraints in Cosmology. This
chapter analyses the expected number density of HI galaxies as a function of
redshift as well as its clustering bias taking into account the flux sensitivities
for HI galaxy surveys using the planned SKA telescopes. We discuss the process by which we obtain these numbers - using a set of state of the art simulations - identify inherent limitations in the current set up and suggest a series of improvements. Although we focus on SKA1 and 2, the sensitivity calculations presented here allow to easily consider other configurations such as an early deployment of SKA1 (corresponding to about 50\% of its expected sensitivity).

\section{Sensitivity calculations}

The noise associated to the flux density (flux per unit frequency) measured by an interferometer can be assumed
to be Gaussian with a rms given approximately by
\begin{equation}
\sigma_S \approx \frac{2 k_B T_{\rm sys}}{A_{\rm eff} \sqrt{2\delta\nu\, t_p}},
\end{equation}
for an array with total effective collecting area $A_{\rm eff}$, frequency resolution
$\delta\nu$ and observation time per pointing $t_p$ ($k_B$ is the Boltzmann constant).
The extra factor of $1/\sqrt{2}$ comes from assuming dual polarised systems. Telescope
sensitivities are sometimes quoted in terms of the "System Equivalent Flux Density":
SEFD $\equiv 2 k_{\rm B} T_{\rm sys}/A_{\rm eff}$ or just "A over T":
$A_{\rm eft}/T_{\rm sys}$. 
The effective collecting area depends on
the efficiency of the system which is expected to be the in the range
of 70 to 80\% for the SKA1 dishes.

Note that the expression above gives the flux density sensitivity per resolution beam (not to be
confused with the dish primary beam or telescope field of view). Moreover, this is 
the sensitivity if the full array is used without constraints on the psf (point spread
function or resolution beam). This is what is usually called the natural array sensitivity. Although angular resolution is not a crucial issue in a HI galaxy survey, some weighting of the visibilities might be required in order to obtain a better behaved psf. In that case, the sensitivity of the telescopes will be further reduced. The required shape and resolution of the psf will depend on the source detection algorithm for HI galaxies which is still a subject of active research. For SKA1, using uniform weighting plus Gaussian tapering of the visibilities can produce "well behaved" Gaussian beams at 10 arc-seconds resolution with flux sensitivities that are 30\% to 80\% worse than what is quoted here (see \citealt{braun13}). However, new source extraction algorithms might be able to deal with more complex psfs (for instance, by working in the visibility space directly) and approach the natural array sensitivity. Therefore in this chapter we decided to quote values using the full array, underlining that in reality the situation might be less optimistic.

The equivalent brightness temperature uncertainty is
\be
\sigma_T= \frac{\sigma_S c^2}{2 k_B \nu^2 (\delta\theta)^2},
\ee
where $\delta\theta$ is the angular resolution of the interferometer.
The total temperature is
\begin{equation}
T_{\rm sys}=T_{\rm inst}+T_{\rm sky}
\end{equation}
with $T_{\rm sky}\approx 60 \left(\frac{300\, {\rm MHz}}{\nu}\right)^{2.55}$ K and
$T_{\rm inst}$ the instrument temperature, which is usually higher than the sky
temperature above 300 MHz. For typical instrument specifications, the 
array noise flux rms can be written as:
\be
\label{rms}
\sigma_S = 260\, {\rm \mu Jy}\left(\frac{T_{\rm sys}}{20\, \rm K}\right)
 \left(\frac{25,000\, \rm m^2}{A_{\rm eff}}\right)
 \left(\frac{0.01 \rm MHz}{\delta\nu}\right)^{1/2}
 \left(\frac{1 \rm h}{t_{\rm p}}\right)^{1/2}. 
\ee

For a given survey area, $S_{\rm area}$ we will need approximately
$S_{\rm area}/(\theta_{\rm B})^2$ pointings where $(\theta_{\rm B})^2$ is the
telescope primary beam (instantaneous field of view) with the full width at half
maximum of the beam, given by (in radians)
\begin{equation}
\theta_{\rm B} \approx 1.22 \lambda/\sqrt{A_{\rm dish}},
\end{equation}
where $A_{\rm dish}$ is the effective area of each dish.
The time per pointing $t_p$ is then related to the total integration time
$t_{\rm tot}$ through
\begin{equation}
t_p=t_{\rm tot} \frac{(\theta_{\rm B})^2}{S_{\rm area}}.
\end{equation}
This will increase the time per pointing at the lowest frequencies (for a fixed $t_{\rm tot}$).
Note however that when doing a survey some overlap of the beams is expected in order
to achieve uniformity on the noise across the sky map. Following the SKA1 imaging
performance memo, we assume that the mosaicked beams will correspond to a size of:
\begin{equation}
\theta_{\rm B}^2 \approx \frac{\pi}{8}\left(\frac{1.3 \lambda}{D}\right)^2\ [{\rm sr}].
\label{effb}
\end{equation}
The situation is slightly more complicated for PAFs (Phased Array Feeds) as below a
certain critical frequency (usually taken as the middle of the band), the beams will
remain constant. Therefore, for a PAF with a certain number of beams, $N_b$, we will
assume that above the critical frequency, the full beam will be $N_b\times {\rm SPF}$,
where SPF is the single pixel feed beam as above (eq. \ref{effb}) and below that critical frequency, it
will remain constant.

\begin{table*}[t]
{\renewcommand{\arraystretch}{1.1} 
\tabcolsep=0.10cm
{\small
\begin{tabular}{|c|c|c|c|c|c|c|c|c|}
\hline
 Telescope & Band [MHz] & z & $T_\mathrm{inst} \, [\mathrm{K}]$ & N$_{\rm dish}$ &
 $D_\mathrm{dish} \, [\mathrm{m}]$  & A$_{\rm eff}\, [\mathrm{m^2}]$ &
 beam $[\mathrm{deg}^2]$ $^{\bf a}$ & Flux rms [$\mu$Jy] $^{\bf b}$\\
\hline
 SKA1-MID & 350 - 1050 & 0.35 - 3.06 & 28 & 190 & 15 & 21,824 & 1.78 & 417\\
 SKA1-MID & 950 - 1760 & (0) - 0.50 & 20 & 190 & 15 & 26,189 & 0.48 & 247\\
\hline
MeerKAT & 580 - 1015 &  0.40 - 1.45 & 29 & 64 & 13.5 & 6,413 & 1.68 & 1015\\
MeerKAT & 900 - 1670 &  (0) - 0.58 & 20 & 64 & 13.5 & 5,955 & 0.64 & 1093\\
\hline
MID + MK & 580 - 1015 & 0.40 - 1.45 & 28.5 & 254 & -- & 28,237 & 1.37 &  328\\
MID + MK & 950 - 1670 & (0) - 0.50 & 20 & 254 & -- & 32,144 & 0.51 & 202\\
\hline
\hline
SKA1-SUR & 350 - 900 $^{\bf c}$ & 0.58 - 3.06 & 50 & 60 & 15.0 & 8,482 & 61 (710) $^{\bf d}$ & 1918\\
SKA1-SUR & 650 - 1670 $^{\bf c}$ & (0) - 1.19 & 30 & 60 & 15.0 & 8,482 & 18 (1300) $^{\bf d}$ & 1151\\
\hline
ASKAP & 700 - 1800  & (0) - 1.03 & 50 & 36 & 12.0 & 3,257 & 30 (1250) $^{\bf d}$ & 4996\\
\hline
SUR + ASKAP & 650 - 1670 $^{\bf e,f}$ & (0) - 1.19 & 30 $^{\bf e}$ & 96 & -- & 11,740 & 18 (1300) $^{\bf d}$ & 832\\
\hline
\hline
SKA2 $^{\bf g}$& 480 - 1290  & 0.1 - 2.0 & 20 & 250 $^{\bf h}$& 50 & 400,000 & 30 & 16\\ 
\hline
\end{tabular}}
}
\caption[x]{Telescope configurations.\\
{\bf Notes:} $^{\bf a}$ This is the primary beam (FoV) calculated at the center of the band unless stated otherwise. It is
proportional to $\lambda^2$ (except for PAFs). For the combined telescopes, the smallest beam of the two
telescopes is used.
$^{\bf b}$ Flux density rms at the centre of the band (or the frequency shown in parenthesis) for a frequency interval of 0.01 MHz and 1 hour integration using Eq.~\protect\ref{rms}. This is the natural array sensitivity. Requirements on the shape of the psf can increase this value by about 50\%.
$^{\bf c}$ Only 500 MHz instantaneous bandwidth.
$^{\bf d}$ PAF beams assumed constant below the critical frequency (shown in parenthesis in
MHz) and going as $1/\nu^2$ above that.
$^{\bf e}$ Assuming that all ASKAP PAFs will be replaced to meet the SKA1-SUR frequency band and
instrument temperature of 30K. 
$^{\bf f}$ Assuming that only band 2 will be initially deployed.
$^{\bf g}$ Values only indicative - can be changed. Both beam and flux rms are assumed constant across the band.
$^{\bf h}$ These should be stations (dense aperture arrays).}
\label{tab:telescopes}
\end{table*}
Following the current baseline design \citep{dewdney2013ska1}, table \ref{tab:telescopes} summarises the
specifications for each telescope, while table \ref{tab:surveys} contains the
specifications assumed for the target HI galaxy surveys. Note that, in order for the
sensitivities to be proportional to $\nu$, we assume that the mosaicking is done at
the highest frequency used for the HI survey (i.e. telescope pointings are packed side
by side at the highest frequency). This way, at lower frequencies, the beams will
overlap, since their size varies as $1/\nu^2$, which will increase the time per
pointing with the corresponding improvement in sensitivity. Of course, for PAFs,
this sensitivity will remain constant below the critical frequency. Table
\ref{tab:surveys} shows that both SKA1 instruments (MeerKAT+MID and ASKAP+SUR) have
comparable performance for a HI galaxy survey at the highest frequency band considered
here (band 2). Note however that this assumes that all ASKAP PAFs will have to be
replaced with new generation receivers at some point. On the other hand, at the
lowest frequencies (band 1), SKA1-SUR seems to perform better.
\begin{table*}[t]
\begin{center}
{\renewcommand{\arraystretch}{1.2} 
\tabcolsep=0.10cm
{\small
\begin{tabular}{|c|c|c|c|c|c|}
\hline
 Telescope & Band [MHz] & beam $[\mathrm{deg}^2]$ & Survey area $[\mathrm{deg}^2]$ &
 $t_p$ [hours] & $S_{\rm rms}^{\rm (ref)}$ [$\mu$Jy]\\
 \hline
SKA1-MID $^{\bf a}$ (Band 1) & 350 - 1050 & 0.88 & 5,000 & 1.76 & 315 \\
SKA1-MID $^{\bf a}$ (Band 2) & 950 - 1760 & 0.88 & 5,000 & 1.76 & 187 \\
\hline
MeerKAT $^{\bf a}$ (Band 1) & 580 - 1015 & 1.06 & 5,000 & 2.13 & 696 \\
MeerKAT $^{\bf a}$ (Band 2) & 900 - 1670 & 1.06 & 5,000 & 2.13 & 750 \\
\hline
SKA1-MID + MeerKAT $^{\bf a}$ (Band 1) & 580 - 1015 & 0.88 & 5,000 & 1.76 & 247 \\
SKA1-MID + MeerKAT $^{\bf a}$ (Band 2) & 950 - 1670 & 0.88 & 5,000 & 1.76 & 152 \\
\hline
\hline
SKA1-SUR $^{\bf b}$ (Band 1) & 350 - 900 & 61 (710) & 5,000 & 122 & 174 \\
SKA1-SUR $^{\bf b}$ (Band 2) & 650 - 1670 & 18 (1300) & 5,000 & 36 & 192 \\
\hline
ASKAP $^{\bf b}$ & 700 - 1800 & 30 (1250) & 5,000 & 6 & 645 \\
\hline
SKA1-SUR + ASKAP $^{\bf b}$ & 700 - 1670 & 18 (1300) & 5,000 & 36 & 140 \\
\hline
\hline
SKA2 $^{\bf c}$ & 480 - 1290 & 30 & 30,000 & 10 & 5.14 \\
\hline
\end{tabular} }
}
\end{center}
\caption[x]{Survey specifications. We assume a total observation time of 10,000 hours. The corresponding flux density rms ($S_{\rm rms}^{\rm (ref)}$) is calculated for a frequency interval of 0.01 MHz using the natural array sensitivity.\\
{\bf Notes:} $^{\bf a}$ Values calculated at the target frequency of 1.0 GHz unless stated otherwise. The beam and time per
pointing ($t_p$) are assumed to change as $(\frac{1.0\, {\rm GHz}}{\nu})^2$ across
the band. The flux density rms is assumed to change as $\frac{\nu}{1.0\, {\rm GHz}}$ across
the band.
$^{\bf b}$ Values calculated at the PAF critical frequency (in parentheses, in MHz). Below
that frequency, values are assumed constant. Above it, the beam and $t_p$ are assumed
to go as $1/\nu^2$, and the flux density rms as $\nu$.
$^{\bf c}$ Indicative values. The beam and flux density rms are assumed constant across the band.}
\label{tab:surveys}
\end{table*}

\section{HI galaxy surveys}


To calculate the HI galaxy number density and bias as a function of flux rms, we used
the SAX-sky simulation\footnote{\texttt{http://s-cubed.physics.ox.ac.uk/s3\_sax}}. 
The SAX-Sky simulation consists of a galaxy catalogue containing the position and
several astrophysical properties for each object in a mock observing cone. It was
produced by \citet{2009ApJ...703.1890O} by adding HI and CO properties to the galaxies
obtained by \citet{2007MNRAS.375....2D} through the post processing of the Millennium
dark matter simulation \citep{2005Natur.435..629S}.

The Millennium simulation is a large dark matter simulation with a box size
of 500 h$^{-1}$ Mpc and a spatial resolution of 5 h$^{-1}$ Kpc. It was run in order to
study the formation of dark matter halos and their evolution with time and therefore
comoving snapshots are available at 64 fixed time steps from redshifts 127 to 0.
The simulation was run using a WMAP1-compatible cosmology and a mass resolution of
$8.6\times 10^8\,h{\rm ^{-1} M_{\odot}}$. The halos were identified as
Friends-of-Friends groups containing more than 20 particles, and merger trees were
built to link each halo with its substructures throughout time.
In \citet{2007MNRAS.375....2D} the simulation was post-processed in order to populate
the halos with simple models of galaxies and used a semi-analytical scheme to evolve
the galaxies properties independently unless a merger occurs. The properties of each
of these galaxies includes the mass of the different constituents, such as the mass
of the cold and hot gas components, as well as the galaxy star formation rate (SFR),
Hubble type, etc. 

The atomic and molecular gas components of each galaxy were then obtained by
\citet{2009ApJ...703.1890O} from the cold gas component
based on properties of nearby regular spiral galaxies. Assuming that the cold gas
of these galaxies resides in flat, approximately symmetric disks, it follows
that the local gas pressure is connected to the molecular gas fraction by a known
physically based prescription, which is also compatible with observations (at least for the more HI massive galaxies). The hydrogen
mass of the cold gas component was assumed to be 0.74\% of the total mass. 
In order to properly emulate the light cone, the SAX-sky simulation used only part of
each snapshot of the Millennium simulation as is described in
\citet{2009ApJ...703.1890O}. Therefore, the boxes at fixed redshifts with HI properties
which can be obtained from the SAX-sky simulation are considerably smaller than 
$500\,h^{-1}\,{\rm Mpc}$ along the line of sight. This can undermine
the estimation of statistical quantities such as the galaxy bias (the number densities,
on the other hand, can be easily obtained from the queries in the s-cubed web site).

\subsection{Galaxy number counts}
For the detection of a galaxy, we required that at least two points on the HI line are
measured, that is, the width of the line has to be larger than twice the assumed
frequency resolution of the survey. The idea is to obtain information on the
typical line double peak expected from HI galaxies due to their rotation. This in
principle will remove any galaxy that is seen ``face on'' since it would show just a
narrow peak. More evolved source detection methods can (and should) be explored, to
avoid spurious detections due to RFI, but we will keep this simple approach for now.
A signal to noise of 10 is then usually required for the detection of a galaxy.
We used the following variables and prescription to detect HI galaxies in the SAX
database:
\begin{itemize}
\item
\texttt{zapparent} -- Apparent redshift (including Doppler correction).
\item
Experiment spectral resolution set to $dV = 2.1(1+\texttt{zapparent})$ at the galaxy rest frame (in Km/s). This corresponds to an observed frequency resolution of 0.01 MHz which is what has been assumed for
the sensitivity calculations.
\item
\texttt{hiwidthpeak} [Km/s] -- Line width between the two horns of the HI-line profile in the galaxy rest frame (already corrected for the galaxy inclination).
\item
Following what has been discussed, we take only galaxies where
$\texttt{hiwidthpeak}/2 > dV$.
\item
\texttt{hiintflux} [Jy Km/s] -- Velocity-integrated line flux of the HI-line (in the observers frame).
\item
For each galaxy the observed sensitivity in terms of flux density (flux per frequency) is then ${\rm flux_d} = \texttt{hiintflux}/\texttt{hiwidthpeak}$. Note that this already factors in the relation between velocity integrated flux and the frequency integrated flux as well as the relation between the velocity line width in the rest frame and the corresponding observed frequency.
\item
Finally, we take only galaxies where
${\rm flux}>N_{\rm cut} S_{\rm rms}/\sqrt{(\texttt{hiwidthpeak}/dV)}$ (we took $N_{\rm cut}=10$ for a "10-sigma"
cut).
\end{itemize}

In order to be as general as possible we did not try to match completely the flux rms
used in the query to the survey specifications given in table \ref{tab:surveys}. Instead we  give results
for different values so that a simple interpolation can be used for different surveys.
In particular, if we decide to use a 5$\sigma$ cut instead of
10$\sigma$, it is just a question of looking for the numbers corresponding to the
flux rms that is half the survey value (since the obtained number counts all assumed a 10-sigma cut).

\begin{figure}[t]
\label{pk}
\begin{center}
\includegraphics[width=0.75\textwidth]{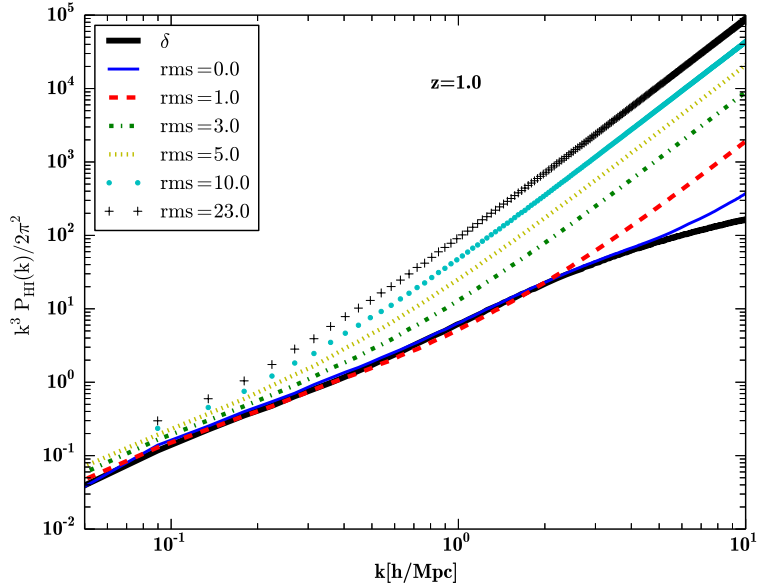}
\caption{The dimensionless HI galaxy power spectrum for different flux cuts using the
prescription described above for each flux rms. Black solid line shows the dark matter
power spectrum from CAMB. The straight lines at high $k$ indicate shot noise.} 
\end{center}
\end{figure}

\subsection{Galaxy bias}
To obtain the bias, the ``detected'' galaxies were put in a box, for which the power
spectrum of the number counts was calculated. The bias squared was then taken as the
ratio of that power spectrum to the dark matter one at $k=0.2$ h/Mpc. 
To perform this study we used boxes with different sizes from $L_\parallel=58 h^{-1}$
Mpc and $L_\perp=100 h^{-1}$ Mpc for redshift 0.06 to $L_\parallel=162 h^{-1}$ Mpc
and  $L_\perp=398 h^{-1}$ Mpc for redshift 2.07.
This makes the bias extraction problematic, since we cannot efficiently probe modes
below $k\sim 0.1$ h/Mpc, where we can safely neglect non-linear effects. For high
flux rms, the number of galaxies is low and the shot noise dominates up to very small
wave numbers. The conclusion is that for high flux/redshift values the results should
be taken as only an indication, in particular for bias above $\sim 20 \mu$Jy flux rms.
As an example we show in Figure \ref{pk} the dimensionless HI galaxy power spectrum at
$z=1$ for different sensitivities. Also note that the Millennium simulation only has galaxies
with masses $\gtrsim 2\times 10^{10}\,M_\odot$ and therefore the bias and number density
for rms=1 might be affected by the lack of smaller galaxies. However this should not
affect the statistics for higher rms values. The cosmological analysis should compare
results for different fiducial values and fully marginalise over the bias and number
counts.


\section{Fitting functions for galaxy number counts and bias}

\begin{figure*}
\begin{center}
\includegraphics[width=0.76\textwidth]{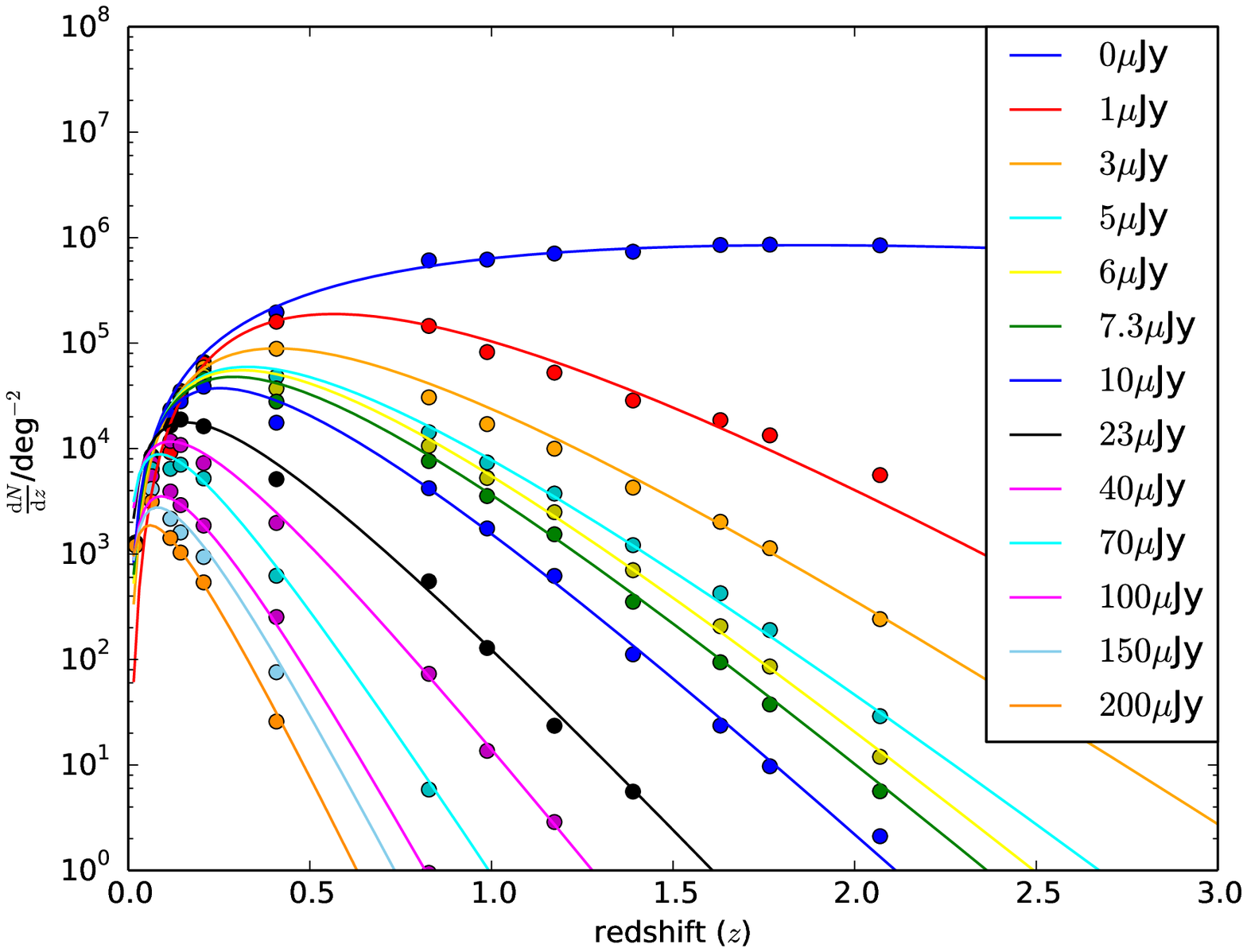}
\includegraphics[width=0.76\textwidth]{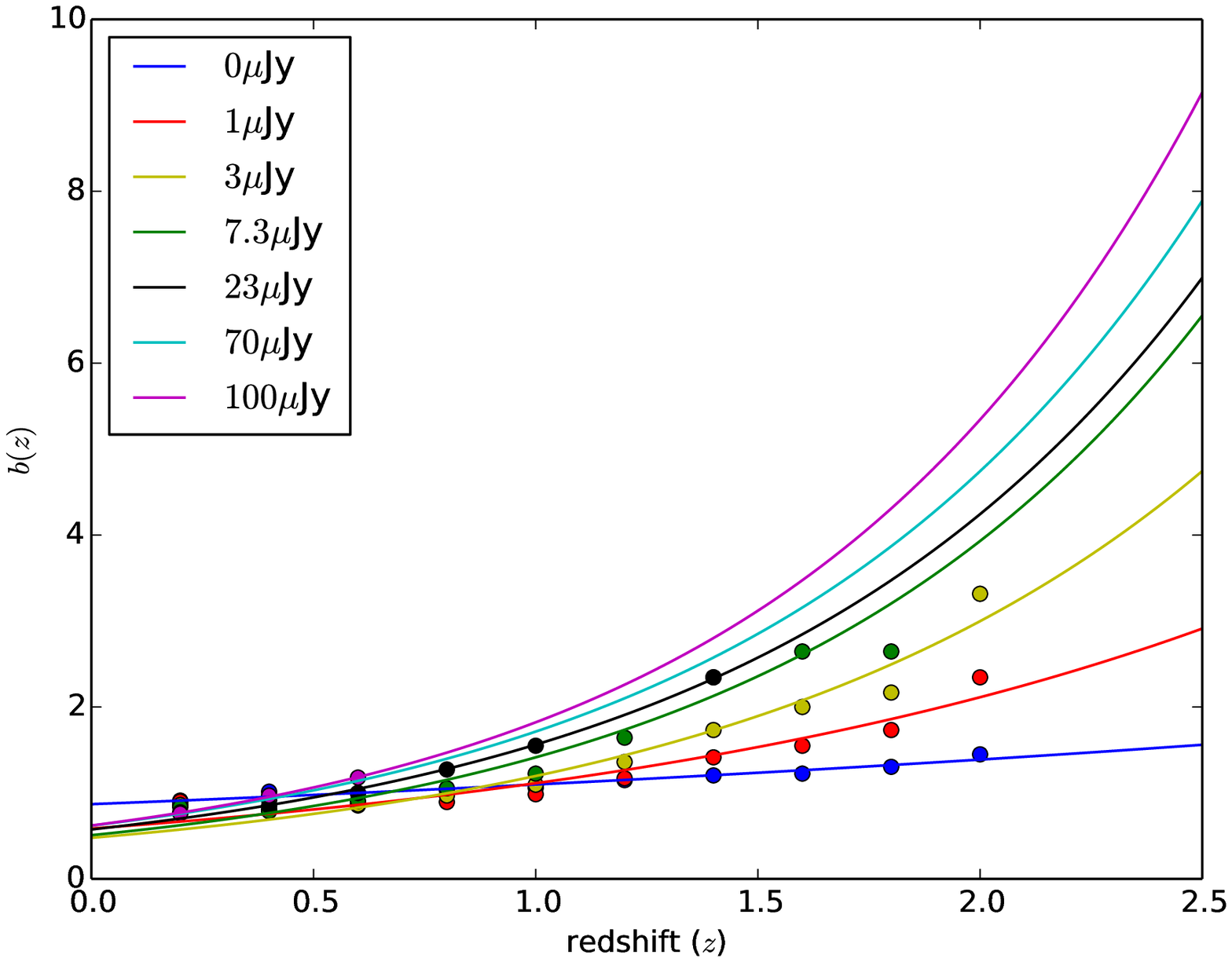}
\caption{Upper panel: Dependence of the HI galaxy redshift distribution $dn/dz$
(units: deg$^{-2}$). Note that the numbers are for different flux rms which will in turn
correspond to a given galaxy flux cut according to the source detection procedure described in the
text (which assumed a 10-sigma cut). Curves follow the fits according to Eq.~\protect \ref{equ: dndz} and dots are
from the S$^3$-SAX simulation. Lower Panel: HI galaxy bias for different
$S_{\rm{rms}}$. Note that above 70 $\mu$Jy values for high redshifts are purely
extrapolations. However, this has little impact as at high z, shot noise will
dominate for these sensitivities.} 
\label{fig:dNOverdz_fit_sax3}
\end{center}
\end{figure*}

In this section we describe fitting functions for the galaxy number counts, $dn/dz$,
and bias, $b(z)$, as a function of redshift. At a constant flux rms, these are well
described by the functions \citep{2014arXiv1412.4700Y}
\begin{eqnarray}
dn/dz &=& 10^{c_1} z^{c_2} {\rm exp}\left( - c_3 z\right) \label{equ: dndz} \\
b(z) &=& c_4 \exp({c_5z}), \label{bias}
\end{eqnarray}
where $c_i$ are free parameters. $dn/dz$ is the number of galaxies per square degree,
per unit redshift. Best-fit values of the parameters are given in
Table~\ref{table:free_parameters}, and Fig.~\ref{fig:dNOverdz_fit_sax3} compares the
fitted curves with data points from the simulations.

Note that for a redshift bin of width $\Delta z$, the observed number of galaxies in
a 1 deg$^2$ patch is $\approx \frac{dn}{dz} \Delta z$, and the comoving galaxy number
density is $n(z)=\frac{dn}{dz}\frac{1}{dV/dz}$, where $\frac{dV}{dz}=\frac{dr}{dz}
(\pi/180)^2 r(z)^2$, $r(z)$ is the coming distance to redshift z and $dr/dz=c/H(z)$.

As explained above, the flux sensitivities for the telescopes listed in Table
\ref{tab:surveys} are taken to either evolve with frequency, $S_\mathrm{rms} \propto
\nu$, or are assumed constant across the band. For the non-PAF receivers (SKA1-MID and
MeerKAT), the flux rms as a function of redshift is
\begin{equation}
S_\mathrm{rms}(z) = S^{\,\mathrm{(ref)}}_\mathrm{rms} 
\left (\frac{N_\mathrm{cut}}{10} \right ) \frac{\nu_{21}}{1~\mathrm{GHz}}
(1+z)^{-1}, \label{eq:Srms}
\end{equation}
where $S^{\mathrm{\,(ref)}}_\mathrm{rms}$ is the reference flux rms at 1 GHz listed
in Table \ref{tab:surveys}, $N_\mathrm{cut}$ is the detection threshold in multiples
of the flux rms, and $\nu_{21} \simeq 1.42$ GHz is the rest frame frequency of the
HI line. For the PAF receivers (SKA1-SUR and ASKAP), the flux rms is
\begin{equation}
S_\mathrm{rms}(z) = S^{\,\mathrm{(ref)}}_\mathrm{rms}
\left (\frac{N_\mathrm{cut}}{10} \right ) \times \left\{ 
  \begin{array}{l l}
    1 & \quad \nu \le \nu_\mathrm{crit} \\[1em]
    \frac{\nu_{21}}{\nu_\mathrm{crit}} (1+z)^{-1} & \quad \nu > \nu_\mathrm{crit}
  \end{array} \right. ,
\end{equation}
where $\nu_\mathrm{crit}$ is the PAF critical frequency listed in Table
\ref{tab:surveys}, and $S^{\mathrm{\,(ref)}}_\mathrm{rms}$ is the reference flux at
the critical frequency. The full SKA has $S_\mathrm{rms}(z) = \mathrm{const}$. We take
$N_\mathrm{cut} = 5$ for all but the full SKA, to which we apply a more
stringent $10 \sigma$ threshold. To obtain the number counts and bias for a given
telescope, we interpolate Eqs. \ref{equ: dndz} and \ref{bias} as a function of flux
rms using the values given in Table \ref{tab:surveys}, and evaluate them as a function
of redshift using the appropriate $S_\mathrm{rms}(z)$ function. We then fit the same
fitting functions to the results -- best-fit parameters are given in Table
\ref{table:flux_scaled_fits}. Note that while the frequency-corrected number counts
are all satisfactorily fitted by Eq. \ref{equ: dndz}, four of the bias functions are
not well described by Eq. \ref{bias}; these are flagged in the table.


\begin{table}
\begin{center}
\begin{tabular}{@{}|c|ccc|cc|}
\hline 
$S_{\rm rms}$ [$ \mu$Jy] & $c_1$ & $c_2$ & $c_3$ & $c_4$ & $c_5$ \\ 
\hline 
 0 & 6.23 & 1.82 & 0.98 & 0.8695 & 0.2338 \\
 1 & 7.33 & 3.02 &  5.34 & 0.5863 & 0.6410 \\
 3 & 6.91 & 2.38 &  5.84 & 0.4780 & 0.9181 \\
 5  & 6.77 & 2.17 & 6.63 & 0.5884 & 0.8076 \\
 6  & 6.84 & 2.23 &  7.13 & 0.5908 & 0.8455 \\
7.3  & 6.76 & 2.14 &  7.36 & 0.5088 & 1.0222 \\
10  &  6.64 & 2.01 &  7.95 & 0.4489 & 1.2069 \\
23   & 6.02  & 1.43 & 9.03 & 0.5751 & 0.9993 \\
40  & 5.74 & 1.22 & 10.58 & 0.5125 & 1.1842 \\
70  &5.62 & 1.11 & 13.03 & 0.6193 & 1.0179 \\
100 &  5.63 & 1.41 & 15.49 & 0.6212 &  1.0759 \\
150  & 5.48 & 1.33 & 16.62 & -- & -- \\
200  & 5.00 &1.04 & 17.52 & -- & -- \\
\hline
\end{tabular} 
\end{center}
\caption{Values of the fitted parameters for Eqs.~\protect \ref{equ: dndz} and
\protect \ref{bias}. For higher flux cuts we suggest taking $b(z)$=1, as only
galaxies below $z\sim 0.5$ are relevant anyway.}
\label{table:free_parameters}
\end{table}


\begin{table}
\begin{center}
\begin{tabular}{@{}|l|ccc|cc|c|}
\hline
{\bf Telescope} & $c_1$ & $c_2$ & $c_3$ & $c_4$ & $c_5$ & $N_\mathrm{gal}$ \\
\hline
SKA1-MID (Band 1) & 4.6399 & 0.8577 & 13.402 & 0.8675* & 0.4767* & $7.40 \times 10^{4}$ \\
SKA1-MID (Band 2) & 5.3662 & 1.2611 & 13.979 & 0.7038* & 0.8616* & $3.39 \times 10^{6}$ \\
\hline
MeerKAT (Band 1) & 4.9849 & 1.8010 & 17.359 & 1.0000 & 0.0000 & $6.27 \times 10^{3}$ \\
MeerKAT (Band 2) & 7.6813 & 4.5740 & 28.067 & 1.0000 & 0.0000 & $1.22 \times 10^{5}$ \\
\hline
SKA1-MID + MeerKAT (Band 1) & 5.1667 & 1.1585 & 14.170 & 0.8116* & 0.7163* & $7.76 \times 10^{4}$ \\
SKA1-MID + MeerKAT (Band 2) & 5.4381 & 1.3315 & 11.836 & 0.6343 & 0.9708 & $4.96 \times 10^{6}$ \\
\hline
SKA1-SUR (Band 1) & 5.1814 & 1.1668 & 14.208 & 0.8108* & 0.7174* & $9.95 \times 10^{3}$ \\
SKA1-SUR (Band 2) & 5.9023 & 1.5879 & 16.149 & 0.6232 & 1.0636 & $4.25 \times 10^{6}$ \\
\hline
ASKAP & 4.8577 & 1.1551 & 17.499 & 1.0000 & 0.0000 & $8.08 \times 10^{5}$ \\
\hline
SKA1-SUR + ASKAP & 5.8611 & 1.4742 & 15.344 & 0.6206 & 1.0559 & $5.55 \times 10^{6}$ \\
\hline
SKA2 & 6.7767 & 2.1757 & 6.6874 & 0.5887 & 0.8130 & $9.12 \times 10^{8}$ \\
\hline
\end{tabular}
\end{center} 
\caption{Values of the best-fit number count/bias parameters (Eqs. ~\protect \ref{equ: dndz} and \protect \ref{bias}) for individual telescopes, after scaling the flux rms with frequency and interpolating. Some of the resulting bias functions are not fit very well by Eq. \protect \ref{bias}; these are flagged. The total number of galaxies expected for each survey is also shown (with 5,000 deg$^2$ except for SKA2).}
\label{table:flux_scaled_fits}
\end{table}

\section{Limitations of current simulations and future improvements}

Although the SAX-Sky is a state-of-the-art simulation in terms of cosmological radio
surveys, it suffers from a number of shortcomings that would be desirable to address
in the future, in order to improve the quality of the forecasts for the SKA.

Probably the most challenging problem for this type of simulations is
finding a compromise between the enormous volume that experiments like the SKA are
expected to probe, and the complexity necessary to simulate the type of objects that
will be observed with sufficient precision. In our case, we would ideally need a
simulation with a box of size $L_{\rm box}\sim15\,h^{-1}{\rm Gpc}$ and resolving haloes
down to $\sim10^9\,h^{-1}M_\odot$, which is simply out of the question given the
present typical computational resources. One must therefore accept a number of
approximations, which simplify the problem while preserving the accuracy of the
relevant physical processes. Recently there has been significant progress, sparked
by the computational needs of the cosmological redshift surveys community, in
finding alternative numerical methods, such as the production of faster simulations
of the matter distribution that are reliable on cosmological scales
(e.g. \citealt{2013JCAP...06..036T,2014MNRAS.437.2594W}).

Even using these methods, it would still be difficult to reach the required mass
resolution. However, a number of methods have been proposed to extend the mass
range in a consistent way \citep{2008ApJ...678..569S,2013MNRAS.435..743D}. Also,
once a simulation has been performed using a particular cosmological model, it can
be efficiently modified to any alternative model in order to study the cosmological
dependence of different observables \citep{2014MNRAS.440.1233M}. This
kind of methods can also be used to generate not just one, but a large number of
realisations, which can be used, for instance to compute uncertainties in a
rigorous way.

Furthermore, while the galaxy population contained in the SAX-Sky simulation was
created applying a semi-analytical model to the dark matter distribution of the
Millennium simulation, it would be of interest (if only for comparison) to use a
prescription based on fundamental principles, using results from hydrodynamical N-body
simulations \citep{2013MNRAS.434.2645D}.

\section{Conclusions}

Simulations calculating the HI content of galaxies will be crucial to predict the number of galaxies that will be observed with future HI surveys. Both the number counts and bias will be fundamental ingredients in order to forecast the constraining power for cosmology of telescopes such as the SKA. In this chapter we analysed explicitly the expected flux sensitivities for the different SKA setups (both for phase 1 and 2) as well as the number counts and bias. Calculations show similar noise rms for SKA1-SUR and SKA1-MID. Numbers were obtained using the SAX-simulations and following a prescription for detecting the HI sources. The bias is shown to have a value around 1 at $z\sim 0.7$, increasing with both redshift and assumed flux cut (dependence on redshift is more prominent for high flux cuts). Assuming a 5-sigma cut, a 5,000 deg$^2$ survey with MID or SUR, should find around $5\times 10^6$ galaxies. This can already have interesting applications for cosmology, but only at low redshifts, $z\sim 0.5$. SKA2 should be able to detect about $10^9$ galaxies over 30,000 deg$^2$ with a 10-sigma cut. This will make it the ultimate "cosmological machine". Further work is still required in order to improve the numbers. In particular: i) source detection algorithms should be compared using real sky and instrument simulations to address the efficiency of HI galaxy detection; ii) larger simulations need to be done in order to account for cosmological scales and in particular provide a better calculation of the bias; iii) full hydrodynamical simulations should be used in order to calibrate the larger simulations and confirm the HI content evolution of galaxies with redshift. Nevertheless, the numbers obtained in this chapter should already give a reasonable forecast of the expected measurements, in particular for the galaxy number counts.

\vspace{0.5cm}

\noindent{\it Acknowledgments:} --- MGS and SY are supported by the South African SKA Project and the National Research Foundation. PB is supported by European Research Council grant StG2010-257080.


\bibliographystyle{apj}
\bibliography{HI_sims}



\end{document}